\documentclass[a4paper,fleqn,usenatbib,onecolumn]{mnras}

\usepackage{newtxtext,newtxmath}

\usepackage[T1]{fontenc}
\usepackage{ae,aecompl}

\usepackage{graphicx}	
\usepackage{amsmath}	
\usepackage{amssymb}	
\usepackage{epsfig}
 
\date{Accepted XXX. Received YYY; in original form ZZZ}

\pubyear{2017}

\title[FRB event rate counts I]{FRB event rate counts I --- Interpreting the Observations}
\author[Macquart \& Ekers]{J.-P. Macquart,$^{1,2}$\thanks{\href{mailto:J.Macquart@curtin.edu.au}{J.Macquart@curtin.edu.au}}  and R.D. Ekers,$^{3,1}$ \\
$^{1}$International Centre for Radio Astronomy Research, Curtin University, GPO Box U1987, Perth, WA 6845, Australia\\
$^{2}$ARC Centre of Excellence for All-Sky Astrophysics (CAASTRO)\\
$^{3}$CSIRO Astronomy and Space Science (CASS), P.O. Box 76, Epping, NSW 1710, Australia\\
}

\begin{document}
\label{firstpage}
\pagerange{\pageref{firstpage}--\pageref{lastpage}}
\maketitle

\begin{abstract}
The fluence distribution of the Fast Radio Burst (FRB) population (the ``source counts'' distribution, $N(>F) \propto F^\alpha$), is a crucial diagnostic of its distance distribution, and hence the progenitor evolutionary history.  We critically reanalyse current estimates of the FRB source counts distribution.  We demonstrate that the Lorimer burst (FRB 010724) is subject to discovery bias, and should be excluded from all statistical studies of the population.  We re-examine the evidence for flat, $\alpha >-1$, source counts estimates based on the ratio of single-beam to multiple-beam detections with the Parkes multi-beam receiver, and show that current data implies only a very weak constraint of $\alpha \lesssim -1.3$.  A maximum likelihood analysis applied to the portion of the Parkes FRB population detected above the observational completeness fluence of 2\,Jy\,ms yields $\alpha =  -2.6_{-1.3}^{+0.7}$.  
Uncertainties in the location of each FRB within the Parkes beam render estimates of the Parkes event rate 
uncertain in both normalising survey area and the estimated post-beam-corrected completeness fluence; this uncertainty needs to be accounted for when comparing the event rate against event rates measured at other telescopes.
\end{abstract}


\section{Introduction}

When a completely new class of astronomical objects is discovered there will be an initial phase when we have no knowledge of their origin and hence no information on their distance.   During this period the analysis of the spatial distribution of the sources over the sky and the distribution of their observed flux densities (i.e. the source counts, or the log $N$--log $S$, curve) provides key information about the distance scale involved.  The spatial isotropy, or the lack of it, indicates whether the objects are local, distributed in the Galaxy, or at great distance, and the source counts will depend on whether or not its distribution is in a bounded volume of space.

Most notable was the discovery of the discrete sources of continuum radio emission in the 1950s.  With only a few radio sources identified with host galaxies it was realised that the steep source counts implied an entire population that was at cosmological distances and had to be evolving in luminosity (or density) with time.  This interpretation was initially controversial as a result of errors in the catalogues \citep{Mills1984} and also due to the tension with the proponents of the Steady State theory \citep{HoyleBurbidge1966,HoyleBurbidge1970}, which predicts no evolution.  However the cosmological interpretation of an evolving population \citep{RyleScheuer1955} prevailed and the complex interplay between the radio luminosity function and the geometry of the Universe involved in source counts was clearly delineated by \cite{vonHoerner1973} (see also \cite{Wall1983} for a more recent detailed analysis).  The evolution required to explain the radio source counts was confirmed when the host galaxies, and in particular the strongly evolving QSO population \citep{Schmidt1968} were identified.

This sequence of events was repeated with the discovery of the Gamma Ray Bursts (GRB) \citep{Klebesadel1973}. \citet{ Paczynski1995} was one of the few to initially argue for an extragalactic origin.  His argument was based only on source counts and on the isotropic spatial distribution.  When the afterglows were eventually detected and the positions were sufficiently accurate to identify the host galaxies the extragalactic hypothesis was confirmed.
 
Fast Radio Bursts (FRBs) are the latest new class of astronomical objects to be discovered, but their origin is still a mystery. Current hypotheses include magnetars \citep{Nicholletal2017}, supramassive neutron stars \citep{FalckeRezzolla2014}, Gamma-Ray Bursts \citep{Zhang2014}, stellar flares \citep{Loebetal2014}, and even the possibility that they may be of artificial origin \citep{LuanGoldreich2014,LingamLoeb2017}.   The slope of the FRB counts can constrain the evolutionary models, providing a test for their different possible origins.  The spatial distribution is nearly isotropic, favouring an extragalactic population, however there is a small but statistically significant avoidance of the Galactic plane \citep{Petroffetal2014}.   As noted by \citet{Vedanthametal2016} the source counts also provides crucial information for the optimum design of future telescopes, if the slope is flatter than $\alpha= -1$ the gain in field-of-view for a smaller dish is more important than the reduction in sensitivity.  

For the FRBs we also have measurements of the pulse sweep rate, and this can be interpreted as cold plasma dispersion in the intergalactic medium if the FRBs are at cosmological distances.  If this assumption is correct and if the intergalactic medium dominates the dispersion we then have a distance indicator.  Note that although one might naively expect fainter FRBs to be more distant and hence to have larger dispersion, it is the shape of the luminosity function that determines whether fainter sources are more distant.  In the case of the radio source catalogs, fainter radio sources are on average closer!  We present the theory connecting the counts of FRBs to their luminosity function and its evolution in a companion paper (Macquart \& Ekers, submitted, hereafter Paper II).  

\citet{Vedanthametal2016} have made a detailed analysis of the FRB counts using both the multiple beam detections and by combining different surveys.  We reconsider their analysis of the multi-beam detections using better Parkes beam measurements and taking into account statistical bias.  We discuss a number of general issues related to methods for estimating the source counts.  In Paper II we will explore the theoretical relationship between the cosmology and the FRB luminosity function and its evolution.  We then relate these to the distribution of dispersion measure for extragalactic models.

The layout of the rest of this paper is as follows. In \S\ref{sec:observations} we discuss the observational constraints on the source counts that can be obtained by the current catalogue of FRBs, and in  \S\ref{sec:issues} we discuss a number of issues which greatly complicate the interpretation of the current observations, including the bias introduced by the discovery process and the effect of the survey telescopes and their beams.  
Our conclusions are presented in \S\ref{sec:conclusions}.

\section{Observational constraints} \label{sec:observations}

\begin{table*}
\caption{Current summary of published Parkes FRB observations from the survey catalogues \citep[][FRBcat]{Petroffetal2016}.  
The centre frequency is 1.35\,GHz in all instances. References are: (1) \citet{Burke-SpolaorBannister2014}; (2) \citet{Keaneetal2011}; (3) \citet{Lorimeretal2007}; (4) \citet{Championetal2016}; (5) \citet{Thorntonetal2013}; (6) \citet{Ravietal2015}; (7) \citet{Petroffetal2015RealTime}; (8) \citet{Petroffetal2017}; (9) \citet{Keaneetal2016}; (10) \citet{Ravietal2016}. $\dagger$The flux density and fluence reported in FRBcat for FRB 150807 are the beam-corrected values; we use the actual measured (i.e.\,uncorrected) fluence of 4.6\,Jy\,ms in our analysis for consistency with the other beam-uncorrected measurements (as explained in the text).
} 
\begin{tabular}{cccccccc}
\hline
FRB &
$l$ (deg) &
$b$ (deg) &
DM (pc\,cm$^{-3}$) &
width (ms) &
$S_{\rm peak}$ (Jy) &
Fluence (Jy\,ms) &
Survey Reference \\
 \hline
010125 & 357 & -20 & 790 & 9.4 & 0.3 & 2.82 &  1  \\
010621 & 25 & -4 & 746 & 7.0 & 0.4 & 2.87 &  2 \\
010724 & 301 & -42 & 375 & 5.0 & $>$30 & $>$150 & 3 \\
090625 & 226 & -60 & 899 & 1.9 & 1.4 & 2.19  & 4 \\
110220 & 51 & -55 & 944 & 5.6 & 1.3 & 7.28  & 5 \\
110626 & 356 & -42 & 723 & 1.4 & 0.4 & 0.56  & 5  \\
110703 & 81 & -59 & 1104 & 4.3 & 0.5 & 2.15  & 5  \\
120127 & 49 & -66 & 553 & 1.1 & 0.5 & 0.55  & 5  \\
121002 & 308 & -26 & 1629 & 5.4 & 0.4 & 2.34 & 4 \\ 
130626 & 8 & 27 & 952 & 2.0 & 0.7 & 1.47  & 4 \\
130628 & 226 & 31 & 470 & 0.6 & 1.9 & 1.22  & 4  \\
130729 & 325 & 55 & 861 & 15.6 & 0.2 & 3.43 & 4 \\
131104 & 261 & -22 & 779 & 2.1 & 1.1 & 2.33  & 6  \\
140514 & 51 & -55 & 563 & 2.8 & 0.47 & 1.32 & 7  \\
150215 & 25 & 5 & 1105 & 2.8 & 0.7 & 1.96 &   8  \\
150418 & 233 & -3 & 776 & 0.8 & 2 & 1.82 & 9 \\
150807$^\dagger$ & 336 & -54 & 266 & 0.3 & 12.2 & 4.6 & 10 \\
\hline
\end{tabular} \label{tab:FRBdata}
\end{table*}

Our analysis is based on a complete sample of FRBs discovered with the Parkes radio telescope and its multi-beam receiver system. This set of FRBs is presented in Table \ref{tab:FRBdata}.  These observations are all taken with a central frequency of 1.35\,GHz.  The receiver sensitivity and field of view is identical for all these observations.  The back-end was improved during this period and this improved the range of DM and polarization information that could be observed but made only small changes to the sensitivity.

\subsection{Effect of surveys with different instruments}

As noted by \citet{Vedanthametal2016}, it is possible to constrain the FRB counts using FRB surveys with different telescopes with a wide range of sensitivity and FoV.  However, as they also remark, the relative normalisation now becomes essential. We can make a significant simplification by only using FRBs discovered with the same observational constraints.  In this case the relative probabilities of finding FRBs with different fluence is independent of the area covered so all observations can be combined without any need to include the field of view or the effective observing time.  FRBs found with different observational selection, such as the Arecibo FRB (121102) or the new Molongolo and ASKAP FRBs cannot be included in this analysis.  At present this reduces the sample by a few FRBs but it avoids the need to normalize the FRB rates, which cannot be done accurately without knowledge of the slope of the FRB counts for reasons discussed in \S\ref{sec:beam}.  Note the source counts are not normalised in this case and we can only estimate the slope.





\subsection{Counting fluence}

The burst duration influences detectability of events, and this affects the completeness of  FRB event rate estimates as a function of flux density in a manner that cannot yet be fully quantified.  However, the fluence is an easily quantifiable property of a transient which is not affected by the time resolution of the observation.  This measure avoids the selection bias on pulse width given the finite instrumental resolution, and is thus preferable over the flux density.  This use of fluence for FRBs is equivalent to the use of integrated rather than peak flux density to avoid bias due to the angular resolution of a radio source catalog. In Paper II we introduce the formalism to investigate the rate counts of FRBs, modelling both the flux density and the fluence. 

\subsection{Survey completeness}

The completeness limit in fluence of the Parkes detections is reported to be 2\,Jy\,ms \citep{KeanePetroff2015}, yet seven of the detected events lie below this threshold.  The S/N of an FRB detection is fundamentally determined by the ratio of received signal power to the telescope noise power received over the same time interval, $t_{\rm FRB}$.   For a fixed FRB fluence the S/N of the detection scales as $t_{\rm FRB}^{-1/2}$, thus affecting the completeness of telescope surveys to longer-duration events.  It is estimated that the Parkes SUPERB survey is complete to FRBs at the 2 Jy\,ms level for an assumed maximum burst duration of $t_{\rm burst}=30\,$ms \citep{KeanePetroff2015}.
For example, for bursts at the 1\,Jy level whose durations exceed $30/2^2 = 7.5\,$ms would be missed.  Given the distribution of burst durations observed so far \citep{KeanePetroff2015,Petroffetal2016}, we thus expect the source counts to be highly incomplete for bursts with peak flux densities below $\approx 1\,$Jy.  The observed event durations vary from 64\,$\mu$s to 10 \,ms, but given the unknown distribution of durations (partly due to this very limitation), the effect of incompleteness is difficult to correct. 

The second factor that affects completeness in a complex way is the radio frequency interference environment (RFI), but this is not discussed extensively in \cite{KeanePetroff2015}.   The FRBs found with the ASKAP telescope at the radio quiet Boolardy site have a negligible number of RFI events \citep{Bannisteretal2017}, while RFI at Parkes is a significant issue with an impact on completeness that is difficult to quantify.  Future observations at radio quiet sites should eliminate this problem.

The $V/V_{\rm max}$ statistic provides another way to address incompleteness, and this is discussed briefly at the end of Section \ref{sec:MaxLikelihood}.



\subsection{Location in the beam}
\label{sec:beam}

FRB detections measure the peak S/N of each burst, not its peak flux density.  The conversion to flux density requires knowledge of the angular position of the burst relative to the telescope pointing position in order to correct for the beam attenuation. For most bursts there is no way to deduce the location of each burst relative to the beam centre, so this correction cannot be applied. The Parkes multiple beam detections using the multibeam receiver can provide some position constraints but, since it under-samples the focal plane, any position solution is degenerate and the corrections may be wrong.  

When including events in the source count distribution it is therefore appropriate to include them at their observationally-measured fluence, rather than using beam-corrected values for some, which will introduce a strong bias because only high S/N events can be corrected.  Any estimate of the source counts requires that all events are treated consistently.  A calculation of the effects of attenuation by the beam response bears this point out (see below and \citet{MacquartJohnston2015} (their eqns.~9 \& 16).  If the underlying count distribution is a power law, but the burst locations are unknown relative to the beam centre, the resulting distribution is still a power law of identical slope whose amplitude is reduced by a factor which reflects the average reduction in the response, averaged over the entire beam shape.  Any estimation of the source count slope therefore requires either that {\it all} events are corrected for the effect of beam attenuation, or that corrections are applied to no events. Hence for the two high S/N FRBs (the Lorimer burst and FRB 150807 reported by \citet{Ravietal2016}) the measured peak S/N and fluence must be used, and not the corrected values.  

\begin{figure}
\centerline{\includegraphics[width=0.5\linewidth]{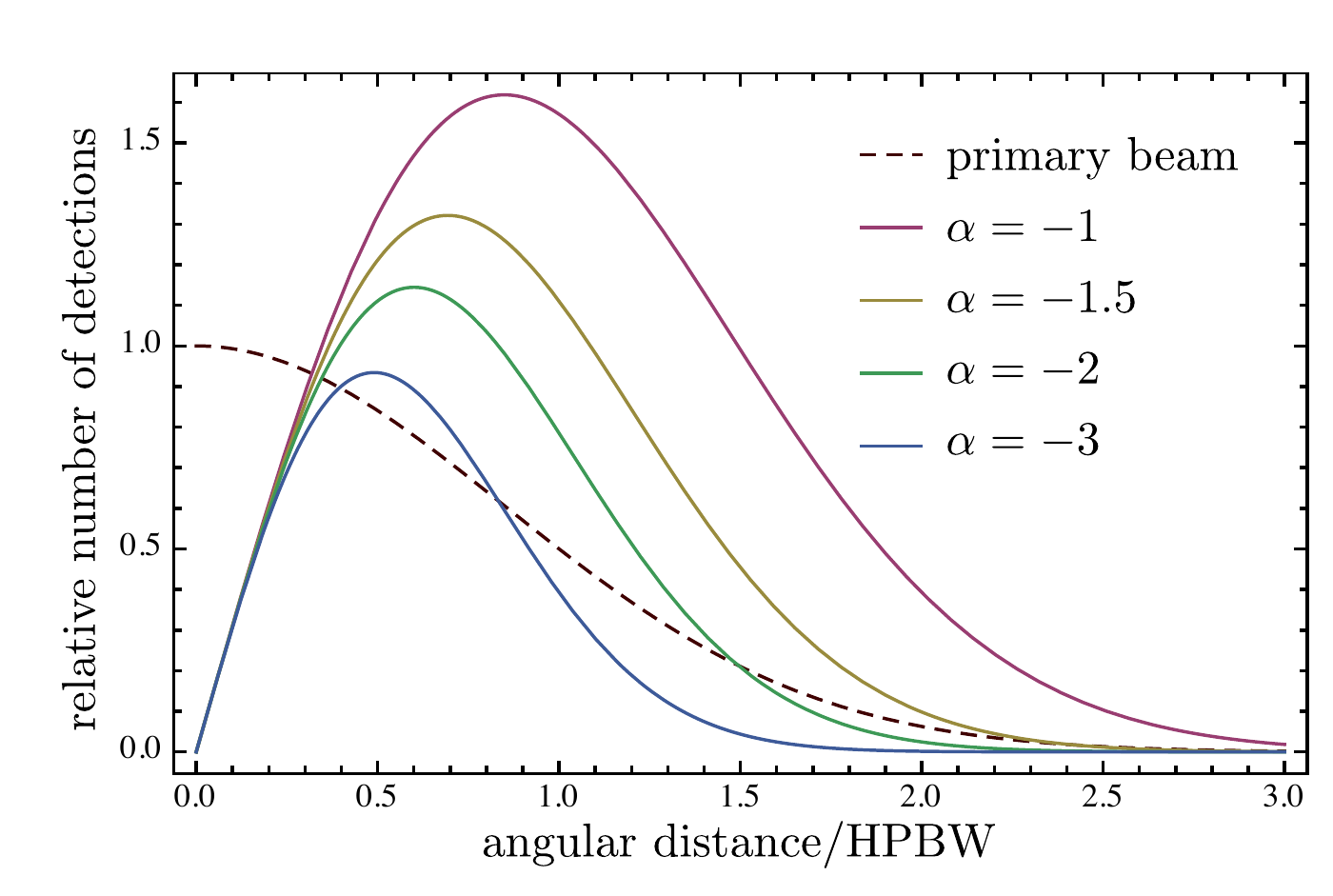}}
\caption{Relative detection rate (arbitrary units) versus position in the beam in units of HPBW for a range of source counts slopes. A Gaussian beam is assumed here.} \label{fig:ProbDetectParkes}
\end{figure}

Although we do not know the FRB location in the beam we can estimate the most probable location and make a statistical correction to the rate.  The area of the beam, and hence the survey area, increases as  we go away from the centre of the beam so the probability of finding a source increases by an amount which depends on the slope of the source counts.  We can quantify this effect precisely in terms of the integral counts per unit area above a flux density, $N(>S_\nu) = K S^{-\alpha}$. If the threshold detection flux density is $S_0$ at the boresight, and the beam power response is a fraction $f_{\rm beam}(\theta)$ relative to that at the pointing centre, the number of events detected in an annulus $\theta$ from the boresight is $2 \pi \theta N(>S_0/f_{\rm beam}) d\theta$.  This is illustrated in Fig.\ref{fig:ProbDetectParkes}, which plots the probability of location for a range of integral count slopes for a gaussian beam.  In this context it is of interest that the deduced position of FRB 150807 is 15$^\prime$ displaced from the pointing centre \citep{Ravietal2016}.  This is twice the HPBW and the correction to the peak flux density applied was more than a factor of 10.  This is not a small effect!

We can use an estimate of the slope of the FRB counts to estimate the most likely location and then make a statistical correction for both the amplitude and the area surveyed.  
For the range of $\alpha$ shown in Fig.\ref{fig:ProbDetectParkes}, the most probable location varies between 0.6 and 1.06 HPBW, the average fluence varies by a factor of 1.7, and the FoV actually surveyed varies by a factor of three.  Hence, from knowledge of the slope of the counts we can already make a very significant correction to the FRB rates.  For a Gaussian beam of HPBW $\theta_b$, the beam is $f_{\rm beam} = \exp[-\theta^2 \log 2/\theta_b^2]$, and the event rate integrated over the entire beam is 
\begin{eqnarray}
R = \frac{\pi \theta_b^2}{\alpha \log 2} K S^{-\alpha}.
\end{eqnarray}
As already noted, this does not change the slope of the counts, but the effective survey area, $\pi \theta_b^2/\alpha \log(2)$, and hence the deduced event rate per unit area, depends strongly on the source counts slope.  This correction needs to be applied to present estimates of the event rate measured with the Parkes telescope.


Future fully-sampled focal-plane arrays (FPAs) will not have this problem.  The FRBs now being discovered with ASKAP using antennas with FPAs have correct positions and hence correct amplitudes and search area \citep{Bannisteretal2017}.  If these rates are combined with other FRB surveys (e.g.~Parkes with the multi-beam) a correction to the relative event rates will be required.  

\section{Source count estimates and their current limitations} \label{sec:issues}

Current estimates place the FRB event rate in the range 2000-5000 events sky$^{-1}$ day$^{-1}$ above a fluence of 2\,Jy\,ms \citep{Championetal2016,KeanePetroff2015}.  \citet{Vedanthametal2016} have made a detailed analysis of the FRB counts and conclude that the count slope is unusually flat (less than Euclidean).  However, there are some major complications to the interpretation of the event rate as a function of fluence, which we now discuss.

\subsection{Lorimer --- the extreme outlier}

The first detected FRB, which we will refer to as the Lorimer burst, was an event with unexpected properties discovered during a search of archival data from a 1.4 GHz pulsar survey of the
Magellanic Clouds using the multibeam receiver on the 64-m Parkes Radio Telescope \citep[FRB 010724;][]{Lorimeretal2007}.  It is an extreme outlier in the source count distribution and its inclusion or non-inclusion makes a big difference to the source counts.  The Lorimer burst saturated the Parkes receiver so its peak flux density had to be estimated from secondary effects and has a large uncertainty, but this alone would not have made it such an outlier.  It was detected in multiple beams so has some properties of the Perytons which are now known to be caused by local RFI but its extreme flux density ratio between beams and the telescope configuration at the time make it almost certainly a member of the FRB population \citep{Petroffetal2015}.  

A more important consideration is discovery bias, which causes the amplitude of a discovery to be inflated.  This is a well known problem in some fields such as epidemiology, e.g. \citet{Ioannidis2008}. Discovery bias is related to a statistical effect known as the \textit {winner's curse}.  Normally this bias results from random noise which raises the amplitude of a real phenomenon above the sensitivity threshold for the measurement, thus enabling the initial discovery. Future measurements will follow a regression toward the mean value of the phenomenon, which will be lower.  The Lorimer burst was two orders of magnitude above the detection threshold \citep{Lorimeretal2007} so noise bias alone would not have played a role in its discovery if it were not for another crucial factor. The Parkes detection software in use at that time performed a statistical test on the one-bit digitised data to examine if it was noise-like and flagged any which contained an excess of 0's or 1's since this could indicate RFI or an instrumental error. The Lorimer burst was so bright that the data contained some blocks of 1's followed by all 0's, and was therefore excised from the beam in which it was strongest (Bailes, personal communication). The Lorimer burst was only discovered because it was strong enough to have also exceeded the detection S/N threshold in the sidelobes of the surrounding beams, so we now have a clear example of discovery bias.

For one-off transients such as the FRBs we also have another interesting difference in the application of discovery bias.  With the exception of the repeating FRB (FRB121102), a given FRB cannot be remeasured, so we have no regression of future measurements towards a mean.  However, the discovery of the Lorimer FRB was also the discovery of a new population and in this case it is the random variability in the population that generates the bias, not just the measurement error. If the Lorimer burst had not been unusually strong it would not have been dectectable in the sidelobes and there would have been no follow-up FRB searches.  It is most likely that no other FRBs would have been discovered. Hence this type of discovery bias is generated by the population variance, not by the detector noise, and this can be a much larger effect.  To illustrate this, we can estimate the probability of finding an outlier as extreme as the Lorimer burst.  If the source counts are Euclidean, the probability of finding any individual event with a fluence as large as the 150\,Jy\,ms estimated lower limit with a telescope capable of detecting events down to 0.55\,Jy\,ms, the lowest fluence event yet reported at Parkes, is 0.020\%\footnote{If we were instead to take the minimum detectable fluence in this dataset to be 2\,Jy\,ms (the Parkes completeness limit) the probability of detection would still be $<0.15\%$.}. The corresponding single-event detection probability even for source counts as shallow as $\alpha=1$ is $<0.4\%$. 


Now we can ask whether this discovery was so unexpected that the threshold for announcing it would be this high.  The answer is clearly yes because the original Lorimer discovery paper \citep{Lorimeretal2007} was treated with considerable skeptism in the community at the time \citep[e.g.][]{Kulkarnietal2014}. That the Lorimer burst fluence exceeds the fluence standard deviation of the rest of the population by 10 times speaks strongly of the magnitude of the sociological (or perhaps psychological) prejudice that nature needs to overcome before we notice a new astronomical phenomenon!  Indeed, the bias against the detection of event so bright was even enshrined in the Parkes detection software, which excluded such strong signals.   


There is a further complication related to the discovery of the Lorimer burst which results in another bias.  The Lorimer team were searching for single pulses (RRATs) in an archival survey for pulsars in the Magellanic Clouds.  Since they were looking for pulses from the Magellanic Clouds the dispersion search space was expanded beyond that normally used for Galactic sources (only 25\,pc\,cm$^{-3}$ in this direction) to 500\,pc\,cm$^{-3}$.  The Lorimer burst would not have been discovered with the normal search parameters since its observed DM was 375\,pc\,cm$^{-3}$, even greater than that measured for pulsars in the Magellanic clouds (70-205\,pc\,cm$^{-3}$). 

The magnitude of these biases is impossible to estimate and for this reason we consider it essential that the Lorimer burst be excluded from any statistical study of the FRB population which involves using its amplitude, fluence or DM.  In this sense we can say the the Lorimer FRB has to be excluded because it suffers from the \textit{winner's curse}.  Although it is excluded in our following analysis, in some cases we note the effect of including it for comparison with other studies.  We note that the entire Parkes data archive, including the Lorimer event, has been re-analysed uniformly using improved search algorithms.  This is important to ensure uniformity in the analysis of the subsequently discovered FRBs, but it is not relevant to the discovery bias issue.



\subsection{Maximum likelihood} \label{sec:MaxLikelihood}

We now address the problem of analysing source counts with a small number of sources.  Integral counts have a special advantage for small numbers because each source is counted at its observed value and no binning is required.  However, it has the disadvantage that the integrated numbers are correlated, so visual inspection may look misleadingly good and the error analysis for any fit is complicated \citep{Crawfordetal1970}.  To avoid the correlation it is common practice to use differential counts with independent bins, but with small numbers the choice of bin sizes and location has a large impact on the results.  As an alternative to both these methods, we have used a maximum likelihood test where the only inputs are the event fluences themselves.  The advantage is that there is no binning involved at all.  The test examines the likelihood that the events in our list could have been drawn from a particular distribution, and we then vary the parameters in the proposed distribution to determine the relative likelihood of the model parameters. 

\begin{figure}
\centerline{\includegraphics[width=0.5\linewidth]{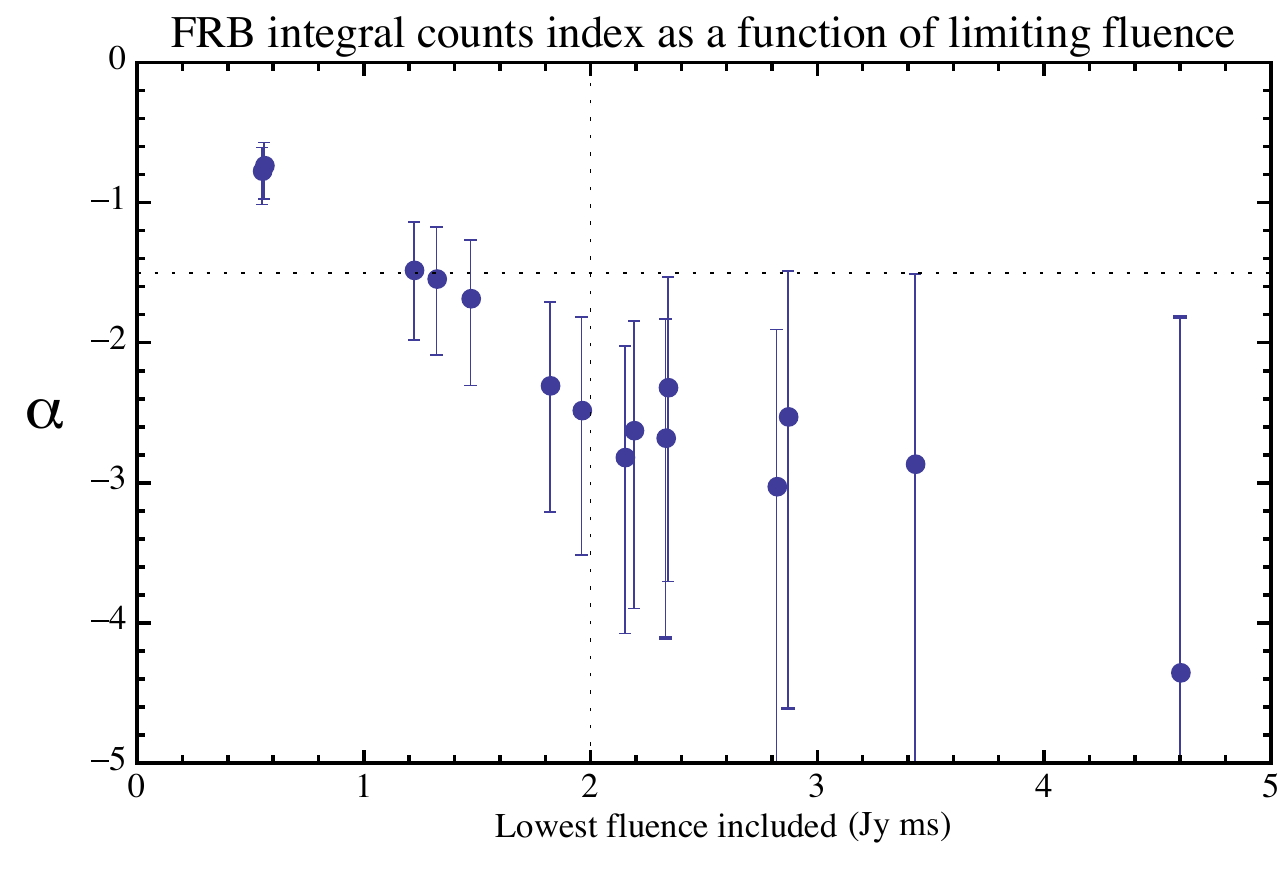}}
\caption{Maximum likelihood source counts slope.  The vertical dotted line denotes the Parkes completeness limit, and the horizontal line indicates the Euclidean source counts index. The error bars denote the 68\% confidence interval of each point, and are derived using the probability distribution of \citet{Crawfordetal1970}, assuming that the trial value of $\alpha$ at that point is equal to the ensemble-average value; as such, they do not take into account the systematic errors associated with incompleteness.} \label{fig:MaxLIndex}
\end{figure}

For integral source counts that follow $N(>F) \propto F^{\alpha}$ above a minimum detectable threshold $F_0$, and where there is no upper bound on $F$ set by the population, \citet{Crawfordetal1970} derive a closed-form solution for the maximum-likelihood value of $\alpha$ and the associated probability distribution of its estimated value.  We have performed a maximum likelihood slope analysis for the subsample of all 16 FRBs found in the Parkes surveys excluding the Lorimer burst, using this solution \citep[eq.\,(9) of][]{Crawfordetal1970}:
\begin{eqnarray}
\frac{1}{-\alpha} = \frac{1}{N} \sum_i f_i,
\end{eqnarray}
where $N$ is the number of sources in the sample and the $f_i=F_i/F_0$ are fluence values, normalised to the minimum detectable value of $F_0$ in the sample.  Figure \ref{fig:MaxLIndex} shows how the estimated power-law slope of the counts changes as we restrict the analysis to a subsample of the events.  Each point in the Figure indicates the value of $\alpha$ as a function of the minimum fluence of that subsample. Below the nominal completeness limit of the Parkes surveys (2\,Jy\,ms) the slope flattens rapidly due to survey incompleteness.  The estimated value of $\alpha$ at $F_\nu = 2.19\,$Jy$\,$ms is $-2.6_{-1.3}^{+0.7}$.
As noted in the previous section, we have not included the Lorimer burst in these estimates.  If it is included, we add a point at very high fluence (far off scale to the right) and the estimated value of source counts index is $\alpha = -1.2_{-0.6}^{+0.4}$ at 2.19\,Jy\,ms. 
If we had included the Lorimer burst and used the beam-corrected fluence of FRB 150807 we would get a maximum likelihood slope of $\alpha = -0.9_{-0.4}^{+0.3}$ at 2.19\,Jy\,ms, which is consistent with the estimated slope of $-0.96 < \alpha < -0.66$ (90\% confidence) obtained by \citet{Vedanthametal2016} .  The different slope estimate in our analysis could be almost entirely explained by the sample definition. 

\begin{figure}
\centerline{\includegraphics[width=0.5\linewidth]{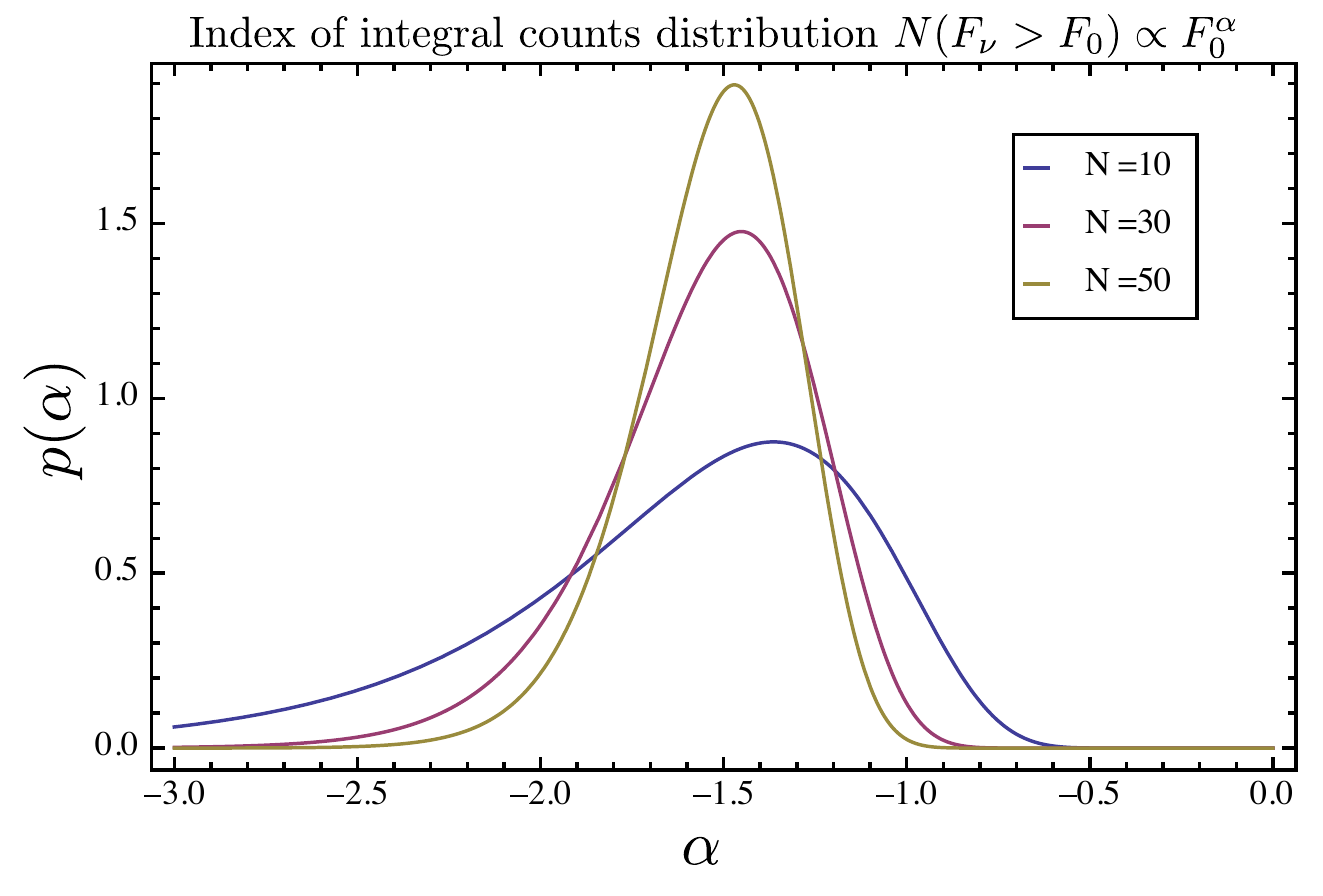}}
\caption{Probability distribution of the estimated integral source counts slope for a population whose ensemble-average index is $\alpha=-1.5$.} \label{fig:MaxLDistn}
\end{figure}

The probability distribution of $\alpha$ \citep[see eq.(12) of][]{Crawfordetal1970} is used to derive the error bars in Fig.\,\ref{fig:MaxLIndex}. This distribution is highly asymmetric about the mean value for small sample sizes, $N \lesssim 30$.  The current sample has only 9 FRBs that can be used above the completeness limit in this analysis, so it is clear that the best estimates of the source count slope will have a large error, and this non-Gaussian error distribution has a long tail extending to steeper slopes. For the subsample comprising only the few stronger FRBs, the errors in this estimate are very large because of the small numbers.  To illustrate the form of the distribution, we plot in Fig.\,\ref{fig:MaxLDistn} the probability distribution of the value of $\alpha$ derived from the method for a range of sample sizes, $N$, for a population whose true source count slope matches the Euclidean value of $\alpha=-1.5$. 

Finally, we remark that an alternative approach discussed elsewhere in the literature \citep{Oppermannetal2016} is to perform a $V/V_{\rm max}$ test, which examines the statistics of the peak flux densities.  
The expectation of the parameter $u=(S_\nu/S_{\rm lim})^{-3/2}$, where $S_{\rm lim}$ is the limiting peak flux density detectable in the survey, yields the value of $\langle u \rangle = \langle V/V_{\rm max} \rangle$.  There is a straightforward relationship between the statistics of $u$ and the source counts distribution.  The expectation of $u$ for a homogeneously distributed population in a Euclidean Universe is $0.5$ and, more generally, for a featureless power-law integral source counts distribution with index $\alpha$, one has $\langle u \rangle = 2 \alpha /(2 \alpha + 3)$.  

The advantage of this test is that the completeness effects related to pulse width are corrected for each FRB, so the test makes near-optimal use of all the available data. However, the applicability of the $V/V_{\rm max}$ test to the published Parkes data is complicated by the inhomogeneity in the time resolution of the searches, and the variety of search packages used to make the detections; the  details of the detection algorithms are not fully encapsulated in the published tables.  Thus we do not feel confident interpreting the $V/V_{\rm max}$ statistics for the published Parkes FRB sample.

\subsection{Multiple beam detections}
As discussed by \citet{Vedanthametal2016}, it is possible to use the number of multiple beam detections in the Parkes multi-beam surveys to estimate the slope of the FRB counts, and this can be done without detailed knowledge of the survey selection thresholds if they are assumed to be the same for all beams.  This is possible because a single survey with a telescope with sidelobes can be considered to be two simultaneous surveys which are made with identical backends, observing time and processing algorithms; one survey has the high sensitivity of the main beam while the other survey has the much lower sensitivity of the sidelobes but with a much larger FoV.  Critical for this analysis is knowledge of (i) the sidelobe pattern of the Parkes telescope and its multibeam receiver, and (ii) the number of events detected in both multiple beams and a single beam. We discuss each of these effects in turn in the two following subsections.

\subsubsection{Parkes beam shape}   

\begin{figure}
\centerline{\includegraphics[width=0.5\linewidth]{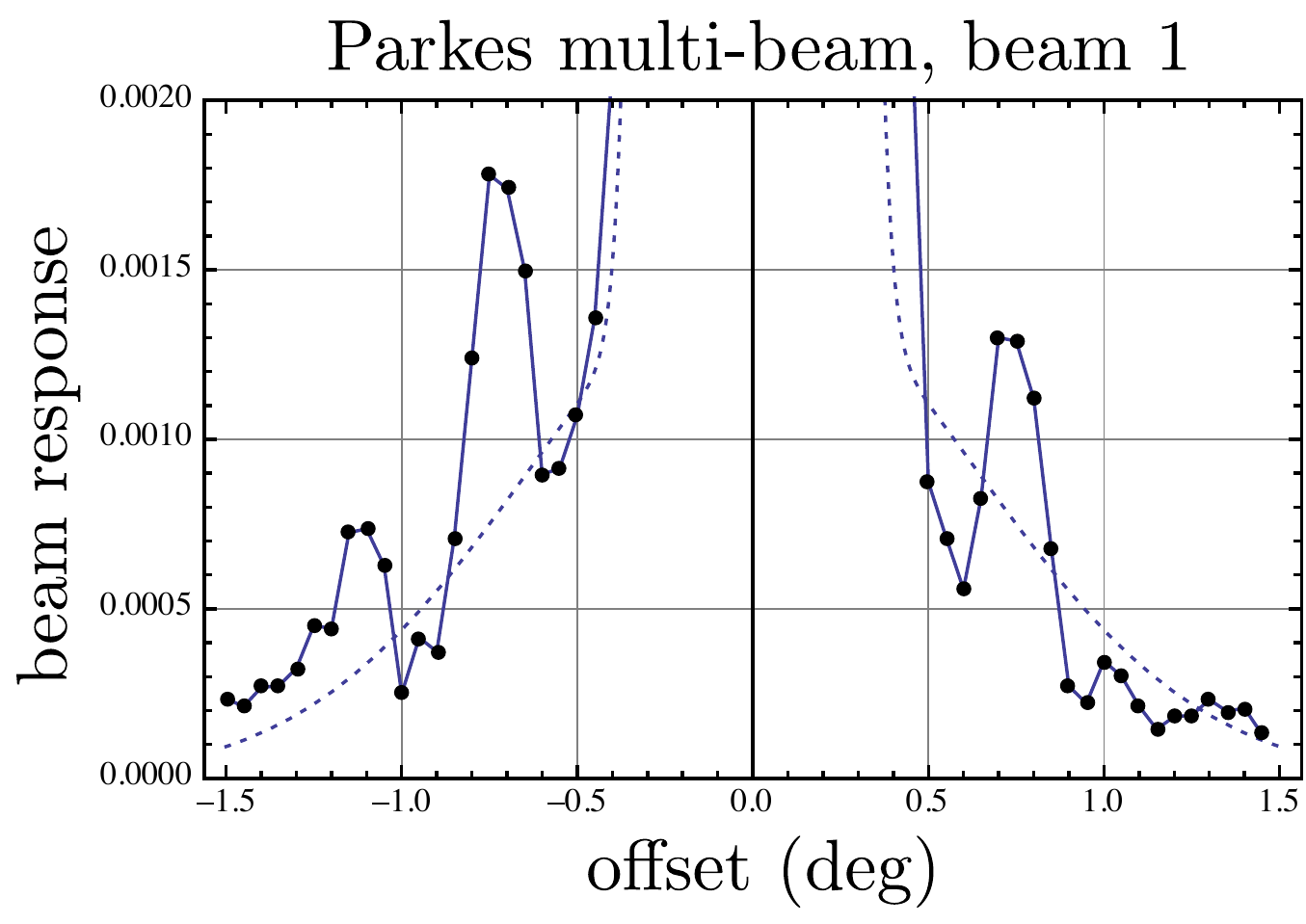}
\includegraphics[width=0.5\linewidth]{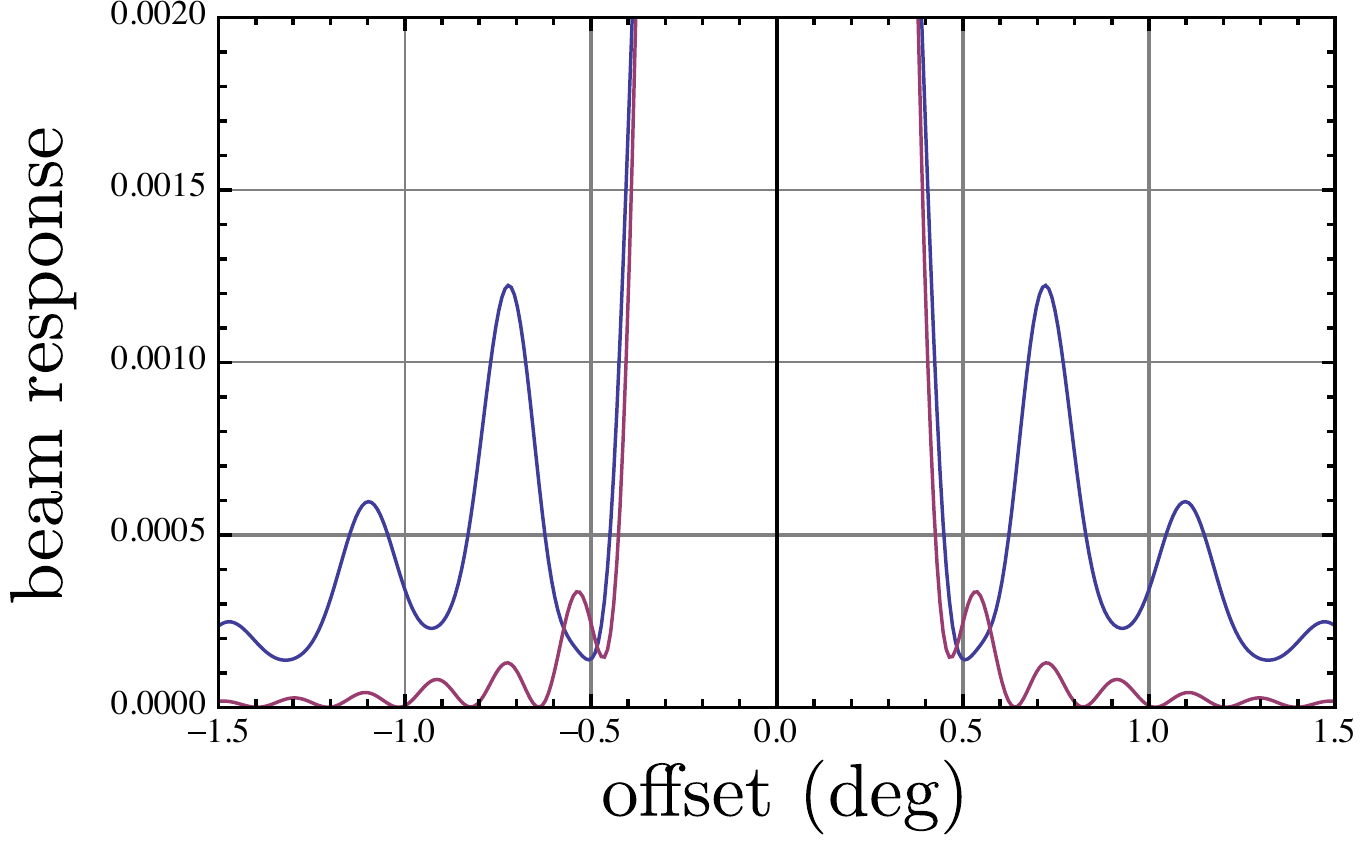}}
\caption{Left: the measured power response of the central beam of the Parkes multibeam receiver compared with the model of eq.(\ref{eqParkesBeam}) superposed (dotted line). The beam shape was obtained using on-off pulse measurements with the Vela pulsar to obtain a clean measurement of the off-source baseline. Right: the result of a recent electromagnetic calculation of the Parkes beam showing the power pattern for a circularly symmetric model, including (blue) and excluding (purple) scattering from the base of the focus cabin (courtesy Alex Dunning).} \label{fig:ParkesBeam}
\end{figure}

\citet{Vedanthametal2016} modelled the Parkes beam patterns using an electromagnetic simulation based on the specifications of the Parkes aperture and the multibeam receiver.  These simulations agreed with the beam patterns published in \citet{Staveley-Smithetal1996}, but since these beam patterns were also based on theoretical simulations with essentially the same parameters, this was only a consistency check.  Unfortunately, neither of these beam patterns agree well with the actual beam measurements, which are difficult at these low levels.  In particular, the electromagnetic simulations miss a significant broad pedestal sidelobe discovered by Staveley-Smith (private communication) from direct measurements of the Parkes beam.  A more accurate recent measurement using the Vela pulsar (George Hobbs, private communication) is shown in Fig.\,\ref{fig:ParkesBeam}.  A pulsed signal is used to measure the low level response and Vela is chosen because it has very narrow scintillation bandwidth and hence low variability across the bandwidth used for these measurements. 

This broad pedestal and its sidelobes are a result of scattering from the base of the 6-m focus cabin.  It has an average amplitude of 0.1\%  and full width of about 1.5 deg (as evident in Fig.\,\ref{fig:ParkesBeam}).  This measurement is for only one scan position angle and is only the centre beam.  Details will change with scan angle and beam offset, so we cannot make a detailed analysis until all these parameters are measured or modelled. However the effect of the focus cabin will be similar for all beams.   Even though this broad pedestal is weak enough to ignore in almost all cases, it has a very large FoV ($100$ times the main beam), and this significantly affects this analysis.  The existence of such a `pedestal' is not new in parabolic dishes, \citep[e.g.][]{Poulton1974}, and is caused by scattering from the conducting base of the focus cabin; in the case of the Parkes telescope this is $\sim 6\,$m, generating the 1.5\,deg pedestal beam. New electromagnetic simulations including this effect (Alex Dunning, personal communication) are in very good agreement with the measured beam, shown in Fig.\,\ref{fig:ParkesBeam}.  The region of the actual feed is well matched and hence acts as an excellent absorber in the centre of the focus cabin. The right panel of Fig.\,\ref{fig:ParkesBeam} shows the large change in the sidelobe structure and amplitude when scattering from the base of the focus cabin are included.  The electromagnetic simulations are based on a circularly symmetric model which is appropriate in our application which integrates the response over the sidelobes.  

These measurements motivate us to estimate the magnitude of the correction due to the additional power using a simple analytic argument.  Consider two adjacent beams centred at positions $\btheta_1=(-\Delta \theta_{\rm sep}/2,0)$ and $\btheta_2 = (\Delta \theta_{\rm sep}/2,0)$ where $\Delta \theta_{\rm sep}$ is the separation between their pointing centres.  The S/N in the $i$th beam ($i=1,2$) of a burst located at some position $\btheta_{\rm B}$ is proportional to the beam response, $f_{\rm beam}( |\btheta_{i} - \btheta_{\rm B}|)$, and if the threshold for detection is $S_0$ when the burst is at the beam centre, then the threshold for detection when the burst is not at the beam centre is given by the condition:
\begin{eqnarray}
S_{\rm B} f_{\rm beam}( \Delta \theta) > S_0, \qquad \Delta \theta = |\btheta_{i}- \btheta_{\rm B}|.
\end{eqnarray}
At a given offset from the beam centre $\Delta \theta$, the minimum detectable burst has a flux density
\begin{eqnarray}
S_{\rm B}(\Delta \theta) = \frac{S_0}{f_{\rm beam}(\Delta \theta)}.
\end{eqnarray}
If the differential source count rate per unit angle is expressed as $dN/dS \propto S^{\alpha-1}$, then the total number detected at a given distance $\Delta \theta$ from the beam centre involves an integration of the source counts from the minimum detectable burst flux density at that point in the beam up to an infinite flux density.  We then integrate over all positions in the beam to derive the total rate:
\begin{eqnarray}
N(\hbox{detected in one beam}) &=& \int_{-\infty}^{\infty} d\Delta \theta_x \int_{-\infty}^{\infty} d\Delta \theta_y \int_{S_B(\Delta \theta)}^\infty dS \frac{dN}{dS} \nonumber \\
&=& C S_0^{\alpha} \int_{-\infty}^{\infty} d\Delta \theta_x \int_{-\infty}^{\infty} d \Delta \theta_y \left[ f_{\rm beam}(\Delta \theta_x,\Delta \theta_y) \right]^{-\alpha},
\end{eqnarray}
where $C$ is a constant.  
We assume for ease of calculation that the beamshape is a monotonically decreasing function\footnote{This simplifying assumption is made so that the beam that the burst is closest to gives the strongest detection. Although, in practice, oscillations in the beam response due to sidelobes violate this assumption, and would require numerical techniques to correctly treat their effect on the source counts slope, they do not alter the fundamental conclusion of the calculation, that there is a substantial change in the estimated source counts slope resulting from an extended beam response of the pedestals.} of $\theta$.   Thus, for the event to be detected in a sidelobe of beam 2 (as opposed to the main beam), the event must be located closer to the centre of beam 1 than beam 2.  This is the case as long as the $x$-component of the burst position is offset by at least a displacement of $+\Delta \theta_{\rm sep}/2$ (and the $y$-component is irrelevant).  Thus 
\begin{eqnarray}
N(\hbox{detected in two beams}) &=& \int_{\Delta \theta_{\rm sep}/2}^{\infty} d\Delta \theta_x \int_{-\infty}^{\infty} d\Delta \theta_y \int_{S_B(\Delta \theta)}^\infty dS \frac{dN}{dS} \nonumber \\
&=& C S_0^{\alpha} \int_{\Delta \theta_{\rm sep}/2}^{\infty} d \Delta \theta_x \int_{-\infty}^{\infty} d \Delta \theta_y \left[ f_{\rm beam}(\Delta \theta_x,\Delta \theta_y) \right]^{-\alpha}.
\end{eqnarray}

We evaluate these expressions explicitly for a Gaussian beam shape, $f_{\rm beam} (\theta) = \exp \left(- \theta^2/2 \theta_b^2 \right)$, where $\theta_b \sqrt{2 \log 2}$ is the HWHP of the beam, and for the Parkes multibeam receiver the beam separation is two FWHM, so that $\Delta \theta_{\rm sep}/2 = 2 \theta_b \sqrt{2 \ln 2}$.  The ratio of bursts detected in both beams to those detected in (at least) one beam is thus
\begin{eqnarray}
{\cal R} = \frac{N(\hbox{two beams})}{N(\hbox{at least one beam})} = \frac{1}{2} {\rm erfc}(2 \sqrt{ -\alpha \ln 2}).
 \end{eqnarray}

\begin{figure}
\centerline{\epsfig{file=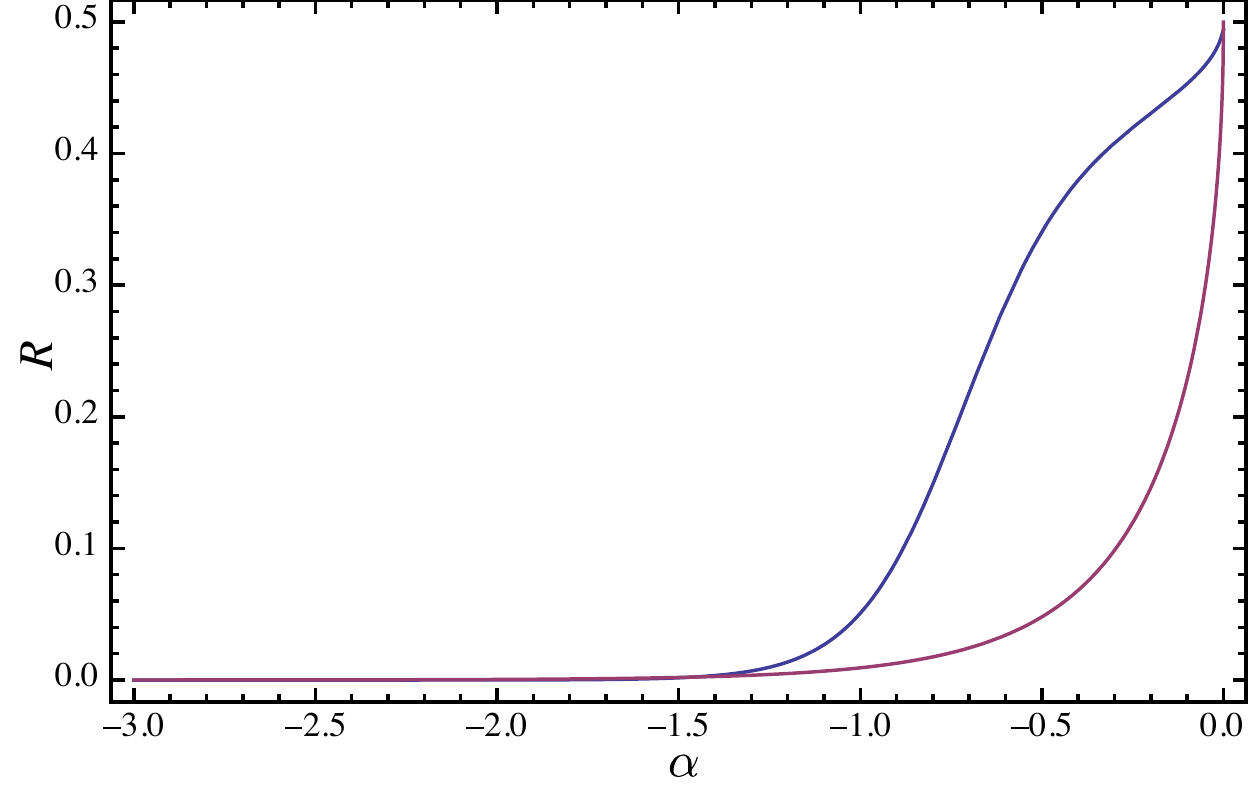,scale=0.5}}
\caption{The ratio of detections in two beams to those detected in (at least) one beam for the Gaussian beam model (purple); and (blue) the Parkes beam shape described in eq.(\ref{eqParkesBeam}) and for the parameters $1-A=0.0015$, $\theta_r=(1.5^\circ/2) / \sqrt{2 \ln 2}$ and $\theta_b=(14^\prime/2)/\sqrt{2 \ln 2}$.} \label{fig:TwoBeamRatio}
\end{figure}

Now consider the additional effect of second, low-power Gaussian contribution to the Parkes off-axis response that extends out to $\approx 1.5^\circ$. We model the beam function with the following form:
\begin{eqnarray}
f_{\rm beam}(\btheta) = A \exp \left[ - \frac{\theta^2}{2 \theta_b^2} \right] + (1-A) \left[ - \frac{\theta^2}{2 \theta_r^2} \right],
\label{eqParkesBeam}
\end{eqnarray}
where $A$ specifies the power in the Gaussian beam relative to the broad low-power Gaussian with FWHM $\theta_r$.  The ratio of multiple beam detection to single (or more) beam detections is 
\begin{eqnarray}
{\cal R}  =\int_{\Delta \theta_{\rm sep}/2}^\infty d\theta_x \int_{-\infty}^\infty d\theta_y f_{\rm beam}(\btheta)^{-\alpha} {\bigg /}
\int_{-\infty}^\infty d\theta_x \int_{-\infty}^\infty d\theta_y f_{\rm beam}(\btheta)^{-\alpha}.
\end{eqnarray}
  
A plot of the ratio, ${\cal R}$, is shown in Figure \ref{fig:TwoBeamRatio} for both a single Gaussian beam shape and the extended beam shape described above in eq.(\ref{eqParkesBeam}).  Parkes beam measurements imply  $1-A=0.0015$ and FWHM values for the main and pedestal beams of $14^\prime$ and $1.5^\circ$ respectively. The effect of the pedestal is most evident over the range $-1 < \alpha <0$.  It is evident that the multiple-beam technique is largely insensitive if the count slope is $\alpha < -1.3$.  


The simple treatment discussed above is not intended to derive a precise revised measurement of $\alpha$, but rather to draw attention to the fact that the estimation of the source counts index using the fraction of dual-beam detections is limited by the detailed shape of the beam response. The solution for the source counts slope is particularly sensitive to the existence of the beam response well off axis, especially if the source counts are flat ($\alpha \gtrsim -1.3$), since this region subtends a large area on sky.  It is difficult to measure this low-level sidelobe pattern for all beams in all directions. This is particularly problematic for the Parkes multi-beam receiver, since it requires detailed knowledge of the beam shapes of all 13 receivers.  This point is exemplified by considering the fraction of detections made out in the sidelobes, beyond the $2 \times $HPBW point, relative to the total number of detections. Numerical integration of the power subtended by the observed beam shape shown in Fig.\,\ref{fig:ParkesBeam}, and assuming circular symmetry, indicates the relative fraction would be 21\%, 8\% and 0.8\% for source count indices of -0.8, -1.0 and -1.5 respectively.  

Of the sample of Parkes FRB detections, \citet{Vedanthametal2016} counted only FRB\,150807 and the Lorimer burst as multiple-beam events.  Excluding the Lorimer burst, the value  ${\cal R}=1/16$ would imply $\alpha=-0.42$ for a simple single-Gaussian beam model, whereas the beam model that includes the effect of the pedestal implies $\alpha=-0.97$.  However, we argue below that FRB\,150807 should not be considered a multiple-beam event.

\subsubsection{Multiple-beam detections}

Having discounted the Lorimer burst, we are left to consider how many events qualify as multiple-beam detections, and to evaluate the number of single-beam detections.  The number of multiple-beam detections is the most crucial issue of the analysis: does FRB150807 constitute a multiple-beam detection according to the criterion of \citet{Vedanthametal2016}? \citet{Ravietal2016} report that its detection in the second beam was at $8 \sigma$, a level at which it would not have been detected in a blind survey at Parkes, whose threshold is $10\sigma$. Its detection in a second beam was possible in this case because the  search is constrained by the parameters derived from the main beam detection. \citet{Vedanthametal2016} analyse the fraction of multiple-beam detections assuming the same detection threshold is applied. Thus, if one were to admit this $8\,\sigma$ event as a multiple-beam detection, one would be forced to also account for all other $8\,\sigma$ events that would be detected by a blind survey at Parkes. 
We note that most of the Parkes FRBs would ultimately satisfy the criterion for multiple-beam detection if their significance thresholds were set sufficiently low.  However, the ratio of genuine (i.e. non-noise) single-beam events to multiple-beam detections should remain the same, since the lowered threshold would then admit a yet-larger number of lower-significance detections which would only be detected in a single beam.  Ultimately, one cannot correct for the number of lower-significance detections without knowing the source counts slope, the very quantity being measured.

On the basis that FRB 150807 can not be classified as a multiple-beam detection for the purposes of the analysis, there are formally no multiple-beam detections, and the ratio ${\cal R}=0$ only provides a weak constraint on the source counts index of $\alpha \lesssim -1.3$.

There is an additional concern related to completeness if we are concerned with the fluence counts.  Since the threshold for detection must fundamentally be expressed in terms of a S/N limit, the issue of incompleteness becomes problematic. Specifically, if the sample is incomplete in fluence at a particular S/N limit, we are forced to estimate the level of incompleteness in our sample, and account for the missed number in the single-beam/dual-beam detection ratio.  Being low fluence events, the missed bursts will predominately be single-beam detections.  For the Parkes completeness limit of 2\,Jy\,ms, the number of missed events can represent a substantial correction. We estimate the magnitude of this correction by estimating the ratio of the integral event rate between 2\,Jy\,ms and the lowest reported Parkes fluence of 0.55\,Jy\,ms relative to the event rate above 2\,Jy\,ms.  This ratio is 6 if the source counts are Euclidean and 27 if the source counts had $\alpha=-2.6$ (the value we estimate from the maximum likelihood method).  Thus, for the 9 events detected above the completeness limit, an extra 47 would have been unaccounted for if the counts were Euclidean.

\subsection{Survey frequency}
The beam location issue also means that the spectral index of the FRBs is not known.   FRBs detected on the side of the main beam will have a steeper observed spectral index due to the beam chromaticity while those discovered in sidelobes could be either steeper or flatter.   At any instant in time the scintillation will also impose transitory spectral structure and the observations of the repeater at both Arecibo and VLA indicates a highly variable spectrum. This will make it difficult to combine the FRB statistics for surveys with telescopes operating at different frequencies.
We note that a recent survey with the CHIME pathfinder reported no detections in a survey of $2.4 \times 10^5$\,deg$^2$h over the frequency range 400-800\,MHz \citep{Amirietal2017}.  However, the interpretation of this rate upper limit on the source counts index of $\alpha > 0.9$, is affected by the foregoing propagation and spectral index issues, complicating the comparison of this number against rates at higher frequencies, particularly the Parkes rate estimate at 1.4\,GHz.

For future FRB searches simultaneous dual-frequency observations (e.g. as provided by shadowing of ASKAP by the Murchison Widefield Array) will provide more readily-interpretable statistics on the frequency structure.

\section{Conclusions} \label{sec:conclusions}


We have shown that a careful analysis of the current FRB population which removes the discovery bias by excluding the Lorimer burst and the beam correction bias on FRB 150807 makes a very large difference to the slope of the FRB counts.  Our best estimate of the slope is $\alpha =  -2.6_{-1.3}^{+0.7}$ at 2.19\,Jy\,ms and we find no evidence for slopes less than Euclidean as had been suggested \citep[e.g.][]{Vedanthametal2016,Calebetal2016}.  However we also note that the error in the estimate of the slope based on the current sample is very large.  We strongly advocate the use of maximum likelihood method for the analysis of data of this type and note the value of using methods which have already been established in the past, to avoid the repetition of old mistakes.

The steeper source counts which are consistent with our new analysis have important consequences for the design of future experiments.  Steeper counts favour systems with greater sensitivity, but the event rate is still sufficiently low that a large FoV is also important.  We note that a steeper power law slope might overpredict the event rate observed using more sensitive radio telescopes such as Aricebo, but there is no expectation that a single power law at high fluences would apply for weaker FRBs.  A lower Arecibo rate would simply indicate that we have reached the expected turnover in the counts.  

One of the biggest issues affecting the interpretation of the current data is the unknown location of the FRBs within the beam when they are discovered, but this source of error will be completely removed with the new generation of survey instruments which fully sample the focal plane.


The paucity of detections at low Galactic latitude can now be revisited \citep{Petroffetal2014}.  The explanation based on scintillation amplification would solve this problem but requires an FRB source count distribution that is significantly steeper than Euclidean, $N(S_\nu) \propto S_\nu^{-2.5}$ \citep{MacquartJohnston2015}. The current analysis favours such a steeper source count slope, so may be consistent with this interpretation.

We have raised many issues relating to the counts of FRBs which indicate that we cannot place much confidence in any estimates of the source counts in the existing literature.  
\section*{Acknowledgements}
Parts of this research were conducted by the Australian Research Council Centre of Excellence for All-sky Astrophysics (CAASTRO), through project number CE110001020.
The authors thank Simon Johnston, Matthew Bailes and Chris Flynn for enlightening conversations on various topics discussed here.  We also acknowledge the use of measurements and models of the Parkes beam provided by George Hobbs and Alex Dunning. 


\bibliographystyle{mnras}
\bibliography{references-FRBCounts1}

\bsp	
\label{lastpage}
\end{document}